\documentclass[letterpaper]{article}

\usepackage[utf8]{inputenc}
\usepackage[T1]{fontenc}

\usepackage{textcomp}
\usepackage{amsmath,amssymb,geometry}
\usepackage{setspace}

\usepackage[style = chem-acs, articletitle=true]{biblatex}
\usepackage{graphicx}
\usepackage{float}
\usepackage{siunitx} 
\newfloat{scheme}{htbp}{los}
\floatname{scheme}{Scheme}
\floatname{chart}{Chart}
\newfloat{graph}{htbp}{loh}


\usepackage{chemformula} 
\usepackage[version = 4]{mhchem} 

\usepackage{authblk}

\author[1,2]{Edoardo Amarotti}
\affil[1]{Chemical Physics and NanoLund, Lund University, P.O. Box 124, 22100 Lund, Sweden}
\affil[2]{ICFO - Institut de Ciencies Fotoniques, The Barcelona Institute of Science and Technology, Castelldefels, Barcelona, 08860 Spain}
\author[1]{Ajeet Kumar}
\author[2]{Luca Bolzonello}
\author[3]{David Bina}
\affil[3]{Biology Centre CAS and Faculty of Science, University of South Bohemia, Czech Republic}
\author[1]{Tõnu Pullerits}
\author[1,*]{Donatas Zigmantas}

\title{A Comparative Study of Coherent and Action-Detected 2D Electronic Spectroscopies of a Multichromophore Photosynthetic System}
\date{*Email: donatas.zigmantas@chemphys.lu.se}

\begin{document}

\maketitle

\begin{abstract}
\noindent
Coherent two-dimensional electronic spectroscopy (C-2DES) is a powerful tool for resolving excitonic structure and dynamics in light-harvesting complexes with dense energy-level manifolds. In contrast, action-detected 2DES has seen limited use on such systems, and its capacity to reveal electronic structure and dynamics remains unclear. Here we compare C-2DES and fluorescence-detected 2DES (F-2DES) on the major plant light-harvesting complex (LHCII) at 78 K. C-2DES resolves excitonic features and their spectral evolution, mapping energy-transfer pathways. F-2DES shows strong cross-peaks, but these are largely time-invariant and dominated by incoherent mixing, obscuring the dynamics. C-2DES is therefore preferred for resolving ultrafast energy transfer in light-harvesting complexes and other large interacting networks. Combined, the two measurements directly quantify the suppression of population dynamics by incoherent mixing, confirming that nearly the entire LHCII trimer network contributes. We also identify pulse-shaper nonlinearities in F-2DES as spurious features, highlighting the need for rigorous artifact suppression.
 
\end{abstract}

\section*{TOC Graphic}
\begin{figure}[h]
    \centering
    \includegraphics[width=0.5\textwidth]{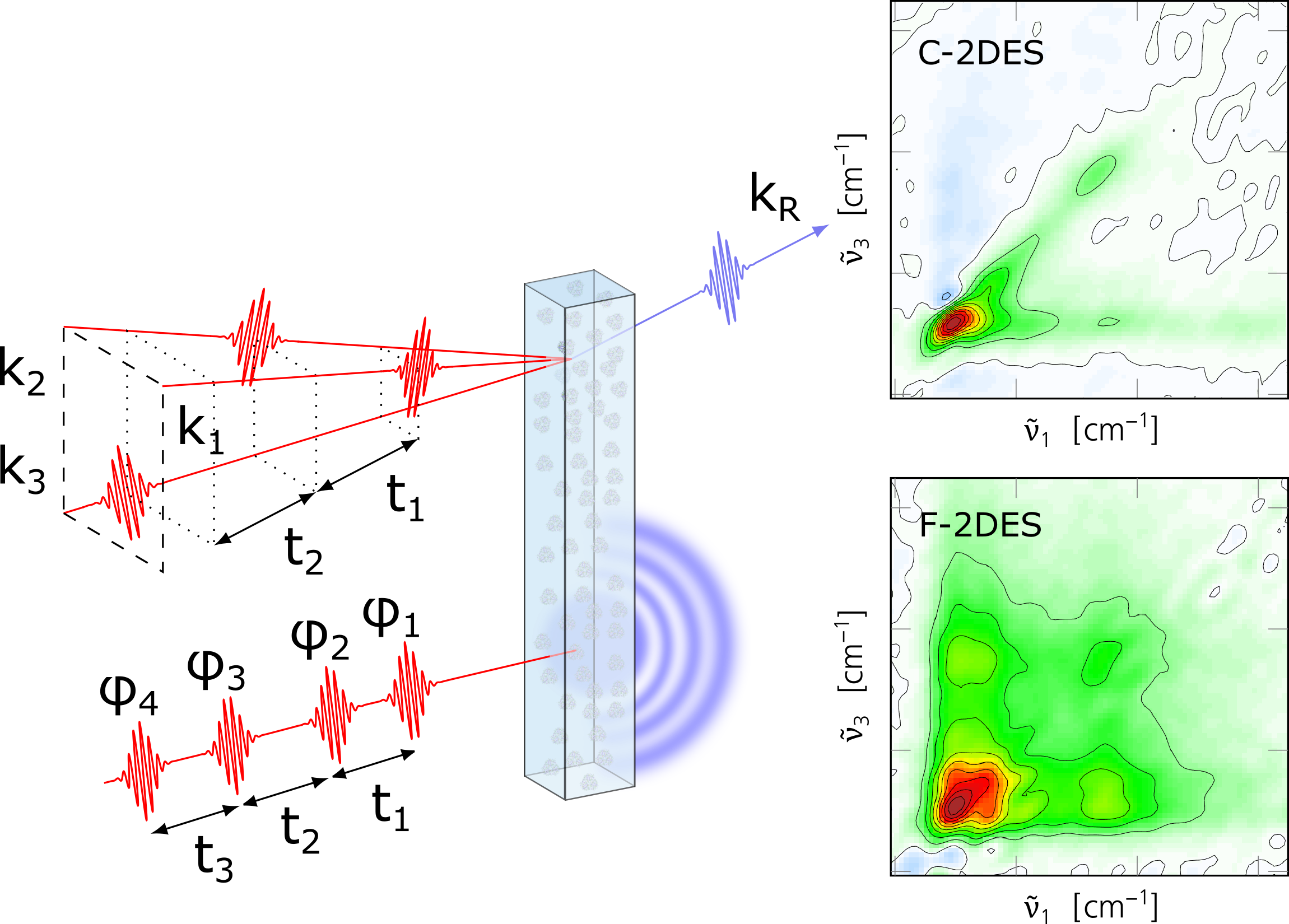}
\end{figure}

%


The solar energy conversion to energy-rich molecules through photosynthesis represents the most fundamental biological process effectively sustaining all life on our planet.
At the core of this process lies the highly-efficient capture of photons by a network of light-harvesting antenna complexes, which funnel the absorbed energy towards the reaction centers often with quantum efficiency approaching unity.\cite{Croce2018,Croce2020}
The light-harvesting complex II (LHCII) stands as the most abundant membrane protein in the biosphere,\cite{Standfuss2005} serving as the main antenna for the photosystem II (PSII) embedded in the thylakoid membranes of green plants and algae.\cite{Caffarri2014,Shen2015,Wei2016,Barber2016} LHCII features multiple electronic transitions, as it binds 8 chlorophyll \textit{a} (Chl \textit{a}), six chlorophyll \textit{b} with four carotenoids and a lipid molecule to complete the system.\cite{Liu2004,Renger2013} 
The primary events in any light-harvesting complex, following photon absorption, are energy transfer steps among the densely packed Chl and carotenoid molecules, which occur on femtosecond to picosecond timescale.\cite{Gradinaru2000,vanOort2010,Lambrev2012,Zigmantas2024} Since the LHCII complexes typically form trimers, excitation can be transferred among Chl molecules of a monomer and  between the different monomers. This forms an interconnected system with intricate energy transfer channels. 
Elucidating the mechanisms governing the energy transfer has long posed a significant scientific challenge, demanding tools capable of resolving the ultrafast dynamics with sufficient temporal and spectral resolution.
Advanced ultrafast spectroscopic techniques have been instrumental in meeting this challenge, providing insights into the photophysics of light harvesting.
Since the 1980s, pump-probe and time-resolved fluorescence spectroscopies helped to unravel main processes in photosynthetic light harvesting.\cite{Zigmantas2024}
Later, coherent multidimensional spectrocopy methods, such as two-dimensional electronic spectroscopy (2DES)\cite{Hybl1998,Fresch2023} was shown to be exceptionally powerful, allowing researchers to map complex excitonic coupling networks, energy flow pathways and their efficiencies.\cite{SchlauCohen2009,Wells2014,Duan2015,Dostal2016,Do2022}
In early 2000s a new type of multidimensional spectroscopy emerged, action-detected 2DES (A-2DES)\cite{Tian2003,Tekavec2006}, which monitors an action observable, typically fluorescence\cite{Tekavec2007,Tiwari2018,Mueller2018,Karki2019,Agathangelou2021} or photocurrent\cite{Nardin2013,Karki2014,Bargigia2022}, which has seen only very limited applications for studies of light-harvesting processes.
Comparisons of the capabilities and potential complementarity of the two techniques are still scarce and were focused on less complex systems.\cite{Maly2020,Javed2024,Ghosh2026}
A central obstacle for the interpretation of A-2DES data is the phenomenon of incoherent mixing taking place during the incoherent detection step, which integrates the signal over the nanosecond fluorescence lifetime.\cite{Schroter2018,Maly2018,Bruschi2023,Bolzonello2023}
This motivates a careful assessment of whether important information survives in complex multichromophoric systems, such as photosynthetic antennae, where incoherent mixing contributions are expected to grow with the number of interacting chromophores and to dominate the response, potentially masking the energy-transfer dynamics that 2DES aims to resolve.
Assessing how severely incoherent mixing affects A-2DES in such systems, and whether any dynamical information survives, is therefore essential for judging the applicability of the method to study highly complex light-harvesting systems.
Here we performed C-2DES and F-2DES in the same core experimental setup on the LHCII complex at cryogenic temperature (78 K).
Analysis of the results enabled us to discuss the differences, limitations and complementarity of the two techniques.



C-2DES experiment was performed in a custom-built setup, the specifics of which have been detailed previously.\cite{Augulis2011}
We used non-collinear optical parametric amplifier (NOPA) pumped by an amplified laser system (Pharos, Light Conversion, Ltd.) which generates sub-20 fs laser pulses.
For this experiment, NOPA was tuned to span the spectral range from 560 nm to 720 nm, with a bandwidth of $\sim90$ nm, which delivers a 13-fs pulse after a compressor consisting of two chirped mirrors and two prisms.
The NOPA output is split into four beams using a plate beamsplitter and a transmission diffraction grating. These beams were arranged in a BOXCAR geometry (in the corners of a square) and focused into the sample.
All parallel linear polarizations $(\langle 0^{\circ}, 0^{\circ}, 0^{\circ}, 0^{\circ} \rangle)$ were maintained for all pulses and signal detection.
To isolate the third-order signal, the first two pulses were modulated by optomechanical choppers, enabling a double-frequency lock-in detection scheme.\cite{Augulis2011}
The signal emitted from the sample was heterodyned with the fourth pulse (local oscillator, with $\sim100$ times attenuated intensity) and the resulting spectral interferogram was recorded by a combination of a spectrograph and CCD camera.

For F-2DES we used a setup based on Dazzler pulse shaper, and the technicalities of the used phase modulation scheme have been described in the previous work.\cite{Amarotti2026}
Briefly, the pulse shaper diffracts the sub-20-fs laser pulse coming from NOPA into four identical collinear replicas, maintaining the spectral shape.
Due to pulse-shaper limitations, only part of the spectrum has been diffracted, resulting in pulses with the spectrum from 620 nm to 710 nm, with a bandwidth of $\sim55$ nm and a pulse duration of 20 fs.
The laser pulses are then focused into the sample via a spherical mirror (radius of curvature 500 mm).
The fluorescence signal is captured and collimated by an off-axis parabolic mirror (f = 100 mm) featuring a 3-mm through-hole that allows the passage of the excitation beam after the sample.
A two-inch-diameter lens (f = 200 mm) focuses the fluorescence onto a pinhole to help filter scattering, after which a long-pass filter (695 nm) rejects residual excitation.
The fluorescence transmitted through the pinhole is re-collimated with a lens (f = 40 mm) and focused onto the high-sensitivity APD (C12703-01, Hamamatsu) with another lens (f = 100 mm).
The APD converts the fluorescence into an electrical signal, which is then sent to a NI cDAQ 9174 voltage reader.
A schematic of the F-2DES setup is shown in Figure \ref{SI-fig:F2DES_Setup} in the Supporting Information.
The fourth-order signal is retrieved by using the phase-modulation scheme described in the previous work.\cite{Amarotti2026}
Despite some differences in their excitation scheme, both C-2DES and F-2DES were implemented to deliver comparable excitation densities to the sample, ensuring a direct and reliable comparison of the sample's nonlinear response.
In the C-2DES measurements, the laser was operated at a repetition rate of 9.4 kHz and focused onto the sample to a spot size of $\sim190$ $\mu$m. This configuration delivered a pulse energy of $\sim2$ nJ for each excitation pulse, which corresponds to an excitation density of $\sim7.1$ $\mu$J/cm$^2$.
For the F-2DES experiments, we employed a 20 kHz repetition rate focused to a tighter $\sim100$ $\mu$m spot size, and pulse energy of $\sim0.65$ nJ for each of the four excitation pulses, and a resulting excitation density of approximately 8.3 $\mu$J/cm$^2$.
Although the focusing geometries and repetition rates differ, the resulting excitation densities are very similar.

\begin{figure}[h]
    \centering
    \includegraphics[]{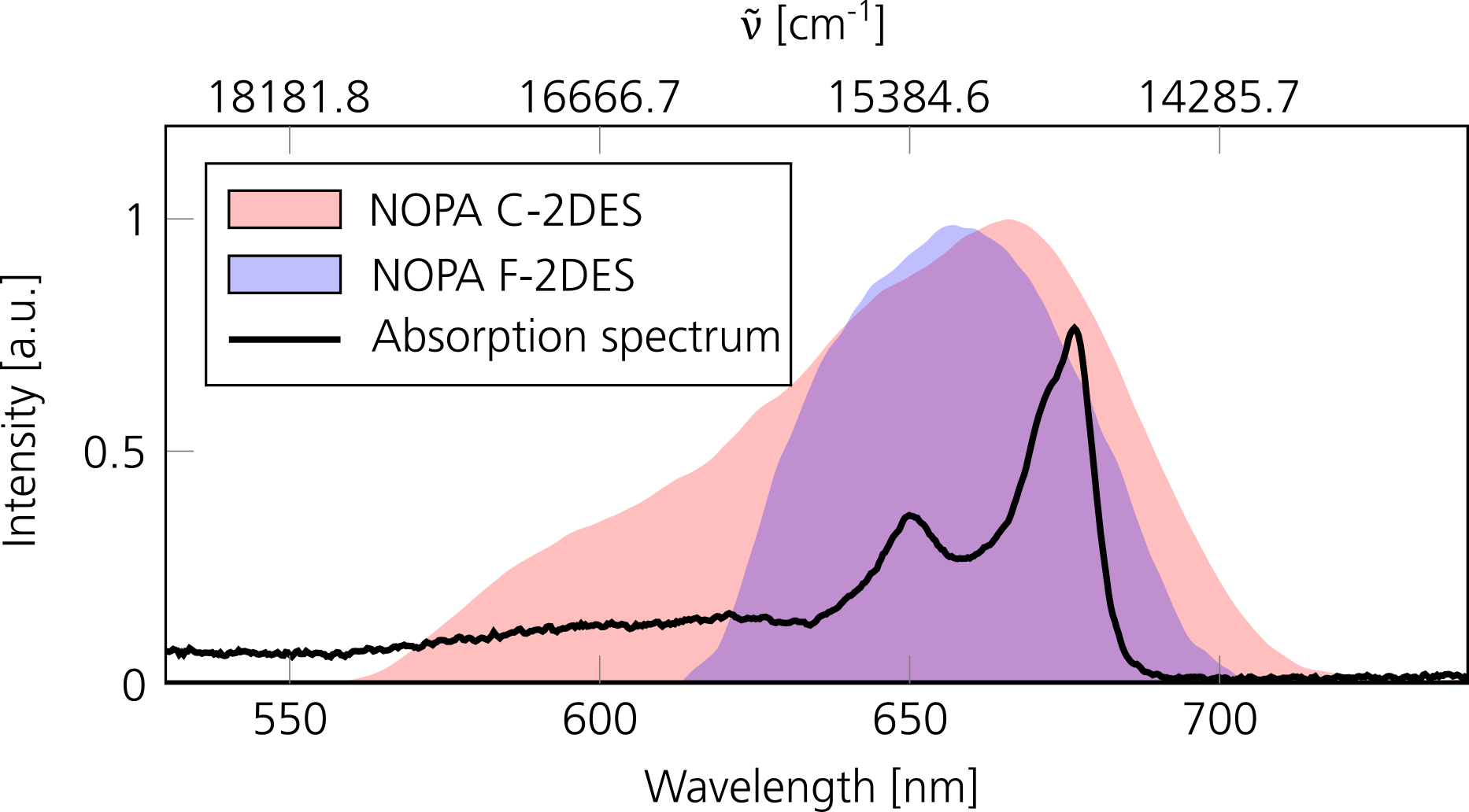}
    \caption{Absorption spectrum of LHCII at 78 K (black line) with the normalized laser spectra used in C-2DES experiments (red shaded area) and in F-2DES experiments (blue shaded area).}
    \label{fig:LinearSpectra}
\end{figure}

Plant LHCII complexes were isolated from spinach leaves, purchased at a local supermarket.
Briefly, the PSII-enriched thylakoid membranes (so-called BBY particles) prepared by Triton X-100 treatment\cite{Berthold1981} were used as the starting material.
BBY particles were stored at $-80$°C prior to further steps.
The LHCII was purified from BBY particles solubilized by n-dodecyl $\beta$-D-maltoside by applying a sucrose gradient centrifugation.\cite{StalevaMusto2019}
The sample was measured in a demountable fused silica cuvette with 200-\textmu m optical pathlength, and it was silanized before the measurements.
The optical density of the sample was kept to 0.14 at $\sim672$ nm at room temperature, as confirmed by the absorption spectrum shown in Figure~\ref{fig:LinearSpectra}.
To achieve a clear optical glass at cryogenic temperatures, the buffer medium was supplemented with glycerol in 40/60\% (v/v) ratio.
After preparation, the sample was immediately inserted into a liquid nitrogen flow cryostat (Oxford Instruments) and kept at 78 K throughout the measurements.

In LHCII the Q\textsubscript{Y} transition, which is the lowest-energy $\pi\rightarrow\pi^*$ electronic transition of Chl, dominates the longer wavelength absorption and fluorescence.\cite{Simonetto1999}
Chl \textit{b} Q\textsubscript{Y} absorbs at $\sim650$ nm and Chl \textit{a} Q\textsubscript{Y} at $\sim675$ nm.\cite{Salverda2003} Figure~\ref{fig:LinearSpectra} displays the absorption spectrum alongside the laser spectral profiles used for C-2DES and F-2DES experiments.
The absorption spectrum is interpreted in terms of coupled chlorophyll pools rather than isolated pigment bands, and at cryogenic temperatures several spectrally-resolved features are visible.\cite{Ramanan2017}
In the Q\textsubscript{Y} region, two dominant bands are evident, which consist of an energy band at 650-660 nm arising from the Chls \textit{b} pools and a band at 670-675 nm arising from Chl \textit{a} pools.
It has been reported that LHCII features at least three different pools of Chl \textit{a} (one trimer and two dimers) and two of Chl \textit{b} (dimers).\cite{Novoderezhkin2005}
The energy transfer from high-energy Chls \textit{b} to Chls \textit{a}, as well as among the Chl \textit{a} pools occurs over multiple timescales, from sub-200 fs to several picoseconds.\cite{Akhtar2019}

\begin{figure}[ht]
    \centering
    \includegraphics[]{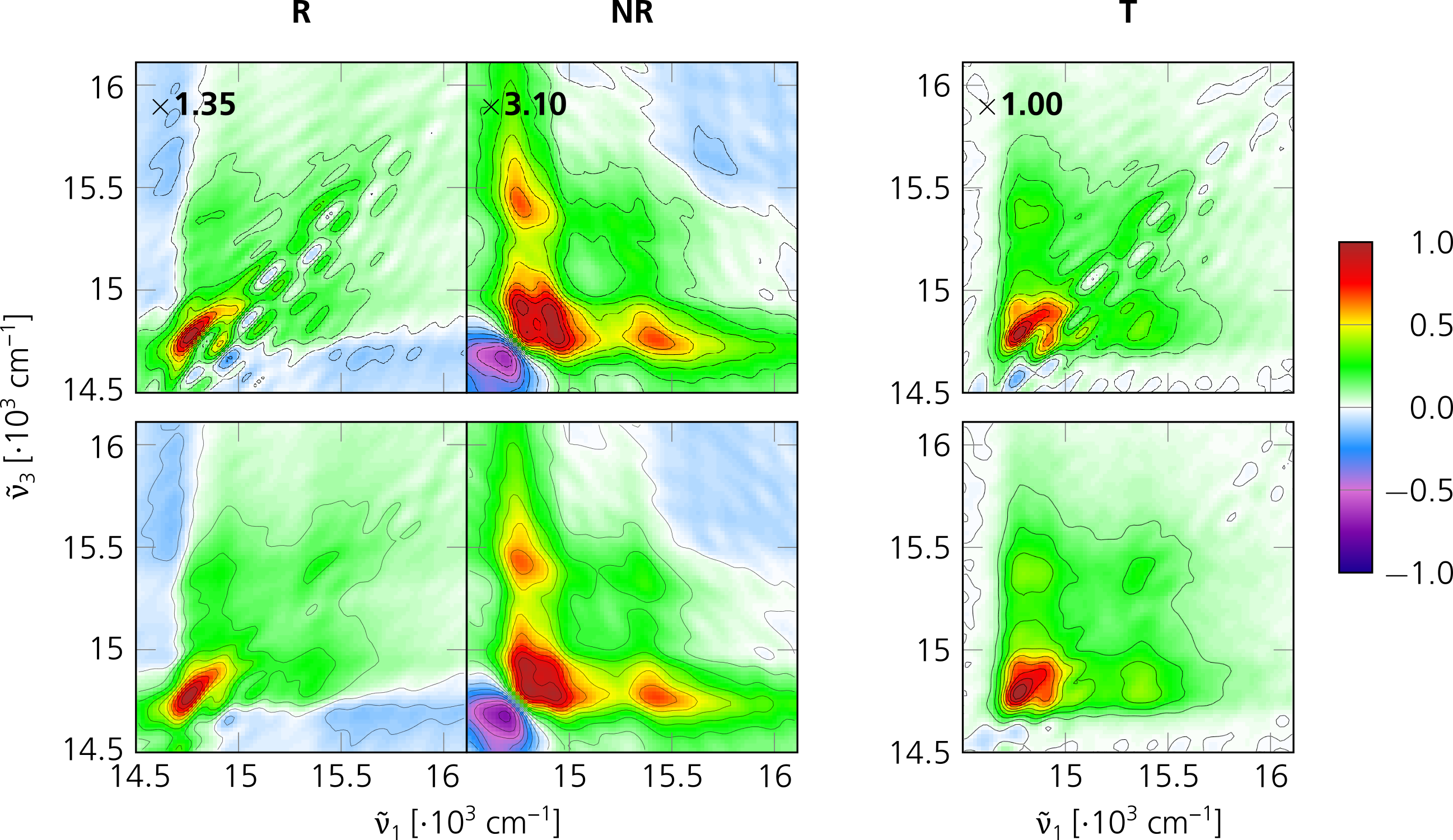}
    \caption{Frequency-domain F-2DES 2D maps of LHCII measured at 78 K at a population time of 200 fs, shown as rephasing (R), non-rephasing (NR), and total/absorptive (T) spectra, the latter obtained by superposition of R and NR. Top row: experimental data; bottom row: artifact free maps after subtraction of scattering measurements. Both R and NR display distortion effects arising from spurious nonlinear components due to limitations of the Dazzler pulse shaper, and these carry over into the total map, which inherits strong distortion features predominantly from the R contribution. The data have been multiplied by the scaling factor shown in the upper-left corner of the panels.}
    \label{fig:FreqTime2Dmaps}
\end{figure}

Now we turn to the F-2DES measurements. In order to obtain the adequate spectral resolution, the coherence time of F-2DES was scanned from 0 to 256 fs.
The corresponding 2D maps in time- and frequency-domain for rephasing (R) and non-rephasing (NR) signals are reported in Figure~\ref{fig:FreqTime2Dmaps}.
The 2D spectra feature a dense set of cross-peaks, a result, which conflicts with the well-established spectral features of the LHCII complex.
This suggests the presence of significant instrumental artifacts, as already observed in other works.\cite{Zhang2025,Amarotti2026}
To extract clean F-2DES datasets, we acquired two nominally identical measurements: one with the sample and the other with the glass cuvette as a scattering medium, keeping all the other conditions unchanged, including the same intensity recorded by APD).
Since scattering from a sample-free cuvette cannot generate a fourth-order nonlinear molecular response, any structured signal observed in this measurement would unambiguously be of instrumental artifact origin.
The scattering measurement, visualized in Figure \ref{SI-fig:RNR2Dmaps_scatter} in the Supporting Information, exhibits a clear nonlinear structure in the time domain.
Upon two-dimensional Fourier transform (2DFT), we obtain spurious features that resemble genuine spectral signals (see Figure~\ref{fig:FreqTime2Dmaps}).
Thus, this test experiment clearly points at the pulse shaper as the origin of these features.
In our data these artifacts are strongest at coherence times t$_{1,3} > 100$ fs; at shorter delays the maps are cleaner (see Figure \ref{SI-fig:FreqTime2Dmaps_cut} in the Supporting Information).
The effect is strongly pathway dependent~---~more pronounced in the R signal, where it produces conspicuous, diagonally aligned peaks, but much weaker in the NR signal (see detailed spectra in Supporting Information).

Our solution to the problem is to additionally measure the time-domain 2D scatter maps and then subtract them from the corresponding sample 2D spectra.
This method is applicable since the features appear to be additive: they show up at regular positions for the same population time and with preserved relative intensities.
However, this subtraction procedure is not perfect, as certain features cannot be completely canceled due to slight mismatches between the datasets.
As a result, small residual artifact features remain, particularly below the diagonal.
Nevertheless, the correction is sufficient to significantly reduce the artifacts, which allows for a reliable analysis of the data.

\begin{figure}
    \centering
    \includegraphics[]{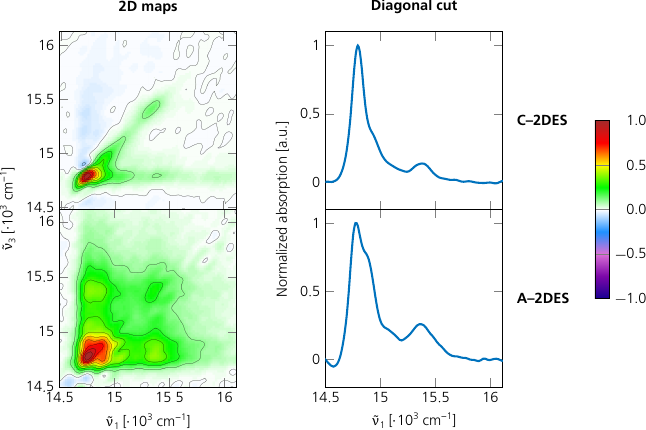}
    \caption{C-2DES (top row) and F-2DES (bottom row) 2D spectra at population time of 200 fs with their corresponding diagonal cuts.}
    \label{fig:C2D_A2D_comparison}
\end{figure}

Now we proceed to the comparison of the C- and F-2DES measurements.
The 2D spectra presented on the left side of Figure~\ref{fig:C2D_A2D_comparison} shows the real (absorptive) total 2D maps for C-2DES (top) and F-2DES (bottom) at population time of 200 fs.
Both 2D maps share many similar fundamental features both along the diagonal, and at the cross-peaks.
As shown in the right panels in Figure~\ref{fig:C2D_A2D_comparison},  diagonal cuts of the 2D spectra from both measurements are similar to the linear absorption spectrum presented in Figure~\ref{fig:LinearSpectra}.

The presence of multiple diagonal and cross-peaks in the C-2DES and F-2DES 2D spectra witness the complex manifold of states and correlations between them, characteristic of the LHCII complex. Specifically, four distinct features can be identified in the $Q_y$ absorption region, consistent with the spectroscopic assignments reported in previous work.\cite{Calhoun2009,Ramanan2017,Do2022,Nguyen2024} The two diagonal peaks at lower energy, located near $14800 \text{ cm}^{-1}$ ($675 \text{ nm}$) are assigned to the Chl \textit{a} manifold, and the higher-energy transitions near $15400 \text{ cm}^{-1}$ ($650 \text{ nm}$) -- to the Chl \textit{b}  states. In both cases, the states in the Chl \textit{a} and \textit{b} regions also include excitonic states of interacting dimers and trimers. The presence of these four features reflects the extended manifold of spectrally distinguishable states that define the fine structure of the complex at cryogenic temperature.
In both measurements, the spectra display well‑defined cross‑peaks that report on the correlations between the distinct electronic/excitonic transitions. These correlations are more clearly visible in the F-2DES measurement.

The differences between the C-2DES and F-2DES  spectra are also evident, arising from the difference in the type of signal that is detected.
As the name states, C-2DES provides a snapshot of the system's coherent response following interaction with three laser pulses. There are three distinct contributions to the signal: excited state absorption (ESA, negative), ground state bleaching (GSB, positive), and stimulated emission (SE, positive).
On the other hand, F-2DES signals are produced following the interaction with four pulses.
Here we also get GSB and SE pathways (both negative), but instead of a single ESA response, two ESA contributions are identified: ESA I (negative), for which the fourth pulse (interaction) brings the system back to the first excited state; and ESA II (positive), where the fourth pulse excites the system to a doubly excited state.\cite{PerdomoOrtiz2012,Kuhn2020}
The doubly-excited state can be a higher electronic state of the same molecule as initially excited, or a second excitation on a different Chl molecule.
In the former case, rapid internal conversion leads to the lowest excited electronic state, which then emits. In the latter case, two excitations of Chls located within the same LHCII trimer can lead to singlet-singlet annihilation (SSA) after which only one excitation remains.
This process is ensured by the efficient energy transfer between the Chls in the LHCII trimer.
Since the detection relies on fluorescence process, which proceeds on a timescale of three to four nanoseconds, there is ample time for the excitations to move around the LHCII trimer and for SSA to occur and consequently influence the signal.
SSA is a crucial nonradiative process in multichromophore complexes in non-linear spectroscopy experiments, and the emergence of ``static'' cross-peaks in action-detected 2D spectra serves as a direct and sensitive probe of its efficiency.
This phenomenon, known as incoherent mixing, has become a research focus in the A-2DES spectroscopy community in recent years.\cite{Gregoire2017,Maly2018,Bruschi2023,Bolzonello2023,Javed2024}

The mechanism behind incoherent mixing lies in the non-perfect cancellation of the ESA II with GSB and ESA I pathways, which occurs whenever SSA takes place: the quantum yield of the ESA II pathway is then diminished and a net cross-peak emerges.
While secondary effects, such as different oscillator strengths between molecules, can also produce imperfect cancellation, annihilation is the dominant mechanism in many coupled systems.\cite{Kalaee2019} This behavior is conveniently captured by a phenomenological factor $\Gamma$, the effective quantum yield of the ESA II pathway, which ranges from $\Gamma = 2$, when the doubly-excited state contributes two units of signal (no annihilation), to $\Gamma = 1$, when annihilation is the exclusive decay route and only one unit of signal remains.\cite{PerdomoOrtiz2012}
Since incoherent mixing can strongly dominate the signal and is static, it poses a challenge for following excitation dynamics in the systems under study (see also below). Several experimental schemes\cite{Faitz2024} and data-analysis methods\cite{Charvatova2025} have been recently proposed to help mitigating the effect.

The severity of incoherent mixing, and the amount of dynamical information that survives it, can be made quantitative within the combinatorial framework of Bolzonello et al.\cite{Bolzonello2023}
For a weakly coupled multichromophoric system in the complete-annihilation limit ($\Gamma = 1$), the cross-peak GSB pathways scale with the product of the transition dipole moments $\mu_i^2\mu_j^2$ of the two involved chromophores, which are symmetric across the diagonal, and remain static on the energy-transfer timescale.
Thus only the SE pathways carry the population dynamics information during t$_2$.
Consequently, the relative weight of the dynamic SE contribution to the static GSB cross-population background is bound by the ratio of the two signals, SE/GSB $\approx 1/N$, where $N$ is the number of optically accessible excitations within the laser bandwidth, rather than the number of physical pigments.

In this work, we excite the full Q\textsubscript{Y} states manifold in LHCII, comprising Chl \textit{a} and Chl \textit{b} electronic and excitonic states, distributed across the three monomers of the trimer. 
Because the inter-monomer energy transfer ($\leq$10~ps) is much faster than the fluorescence lifetime ($\sim$3--4~ns) and most of 42 chlorophylls are connected by the energy transfer pathways, the entire trimer effectively contributes to the detected signal, yielding an effective $N \approx 18$--$24$, once excitonic delocalization within each monomer is taken into account.\cite{Salverda2003,Renger2013}
The SE/GSB ratio is therefore expected to be $\lesssim 5\%$ for LHCII, with a correspondingly small dynamic SE contribution.

Because the GSB cross-population background does not evolve during t$_2$, any population dynamics manifest as a small time-dependent signal on the top of this background, with relative amplitude of $\sim 1/N$.
For LHCII this amounts to only a few percent of the total cross-peak amplitude, i.e. at or below the noise level of the current F-2DES measurement. Since the time-dependent signal cannot be reliably extracted, the estimate of $N$ from F-2DES alone cannot be made.
The C-2DES data removes this ambiguity by providing a mixing-free reference for the dynamics expected at each peak: thus comparing the data from two experiments at the same peak positions both confirms that the dynamics signals are suppressed in the F-2DES measurement rather than completely absent, and quantifies the suppression, which directly yields $N$.

\begin{figure}[ht]
    \centering
    \includegraphics[]{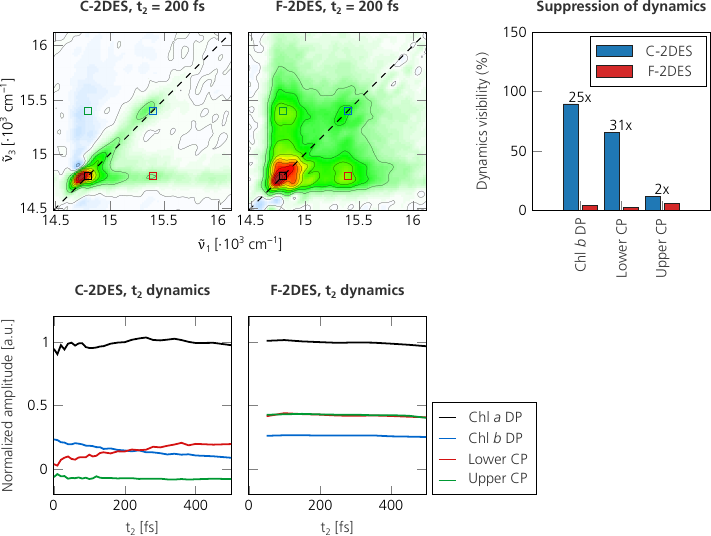}
    \caption{Quantitative comparison of C-2DES and F-2DES dynamics for LHCII.
    \textbf{Top left, top centre:} real absorptive 2D maps at t$_2 = 200~\text{fs}$; 
    colored squares mark the areas ($80~\text{cm}^{-1}$) used for the analysis: chosen Chl~$a$ DP (black), Chl~$b$ DP (blue), corresponding lower CP (red), upper CP (green). \textbf{Bottom left, bottom centre:} corresponding t$_2$ traces, each normalized to the Chl~$a$ DP amplitude at t$_2 = 200~\text{fs}$.
    The Chl~$b \to$ Chl~$a$ transfer signatures clearly visible in C-2DES are strongly suppressed in F-2DES.
    \textbf{Top right:} dynamics visibility factor (see text for details ) for the three peaks, with the C-2DES/F-2DES ratio annotated above each pair.}
    \label{fig:DynamicsComparison}
\end{figure}

This prediction is borne out directly from the side-by-side comparison of the results obtained by the two experimental schemes, as shown in Figure~\ref{fig:DynamicsComparison}.
In C-2DES, the lower cross-peaks (CPs) are substantially more intense than the upper CPs, as expected at 78~K, where the downhill energy transfer dominates.
As a concrete example, we consider one of the two Chl \textit{a} and one of the two  Chl \textit{b} peaks, and corresponding cross-peaks, as marked by colored rectangles in the 2D spectra presented in Figure~\ref{fig:DynamicsComparison}.
The t$_2$ traces accentuates this with the one of the Chl \textit{b} diagonal peak (DP) decaying as the lower CP corresponding to the Chl \textit{b} to Chl \textit{a} energy transfer rises.
In F-2DES, by contrast, all the CPs, symmetric in respect to the diagonal, are essentially equal in amplitude.
Here, the kinetic traces at chosen diagonal and cross-peaks are flat over 0--500~fs time window.

To characterize, in a model-independent way, the suppression of population dynamics by incoherent mixing, we compare the population dynamics extracted from C-2DES and F-2DES measurements at the same peak positions.
For each peak, a region of interest (ROI)-averaged trace $S(\text{t}_2)$ is obtained over a square region of $80~\text{cm}^{-1}$ centered on the assigned excitation and detection energy coordinates, and is normalized by the Chl~$a$ diagonal amplitude measured at t$_2 = 200~\text{fs}$ in its own dataset, placing the two experiments on a comparable scale.
We then quantify the dynamic range of each trace through a dimensionless visibility (V),

\begin{equation}
    V = \frac{\left| S(\text{t}_2^{\max}) - S(\text{t}_2^{\min}) \right|}
             {\left\langle S(\text{t}_2) \right\rangle_{[200, 500]}},
    \label{eq:visibility}
\end{equation}

with t$_2^{\min} = 50~\text{fs}$ and t$_2^{\max} = 500~\text{fs}$.
The denominator ${\left\langle S(\text{t}_2) \right\rangle_{[200, 500]}}$ is the time-average of the trace over the [200, 500] fs window within the square ROI considered, chosen to exclude the early-time region where residual coherent artifacts and shaper-induced features are present (see more details in the Supporting Information).
It serves as the reference amplitude against which the dynamic variation in the numerator is normalized.
Being defined as a ratio, $V$ is invariant under uniform rescaling of $S(\text{t}_2)$ and is therefore a dimensionless measure of the relative dynamic range that can be compared directly between C-2DES and F-2DES, despite their different signal units and absolute amplitudes.
Because in C-2DES the cross-peak carries the full population dynamics whereas in F-2DES only the $\sim 1/N_{\text{eff}}$ SE fraction does, the F-2DES dynamic range is suppressed by exactly this factor; the ratio of visibilities therefore returns $N_{\text{eff}}$ directly.
The ratio $V_C / V_A$ for a given peak then reports the suppression of the population-transfer signature in F-2DES ($V_A$) relative to C-2DES ($V_C$).
Within the incoherent-mixing framework, where the relative weight of SE to GSB contributions at the CP scales as $\mathrm{SE}/(\mathrm{SE} + \mathrm{GSB}_{\text{cross}}) \sim 1/N_{\text{eff}}$, this suppression ratio is itself a direct, model-independent measure of $N_{\text{eff}}$, i.e. $V_C/V_A \approx N_{\text{eff}}$.

As pointed above, we estimate $N_{\text{eff}}$ from the two prominent spectral signatures, the rise of the lower Chl~$b$--Chl~$a$ CP and decay of the Chl~$b$ DP (see the top-right panel in Figure~\ref{fig:DynamicsComparison}), which yields $N_{\text{eff}} \approx 28$, of the same order as the $\approx 18$--$24$ optically accessible transitions expected for the interconnected trimer.
Following Javed \emph{et al.} \cite{Javed2024}, two effects bear on this comparison and act in opposite directions.
On one hand, excitonic delocalization redistributes oscillator strength and pushes the SE/GSB ratio above the combinatorial $1/N$ limit, which by itself would \emph{lower} the apparent $N_{\text{eff}}$ relative to the bare transition count.
On the other hand, the finite F-2DES laser bandwidth and the slight spectral overlap of Chl~$a$ and Chl~$b$ bands reduce the apparent dynamic contrast $V_A$ and thereby \emph{raise} the apparent $N_{\text{eff}}$.
In our measurement, this latter effect potentially dominates and accounts for $N_{\text{eff}} \approx 28$ exceeding the structural estimate, with the precision ultimately limited by the small, noise-limited $V_A$.
We therefore interpret $N_{\text{eff}} \approx 28$ as a model-free confirmation that the incoherent-mixing pool comprises essentially the full set of electronic transitions in the LHCII trimer, while cautioning that its value reflects the balance of these competing effects, the noise level in the F-2DES measurement, and should not be read as an exact transition count.  

The suppression measured at the two peaks does not report on a single global number, but rather on two distinct detection manifolds. Following the combinatorial counting of Bolzonello \emph{et al.}\cite{Bolzonello2023} and Javed \emph{et al.}\cite{Javed2024}, the SE/GSB ratio at a given peak is set by the pool of chromophores probed by the second interaction pair: the Chl~$b$ DP reports on the Chl~$b$ detection manifold (18 Chl~$b$ in the trimer), whereas the lower Chl~$b$--Chl~$a$ CP reports on the Chl~$a$ acceptor manifold (24 Chl~$a$). In the site basis these correspond to expected suppressions of $\approx 18$ and $\approx 24$, respectively, and excitonic delocalization can only reduce these values. Experimentally we obtain $V_C/V_A \approx 25$ at the Chl~$b$ DP and $\approx 31$ at the lower CP: of the correct order, and with the correct ordering (larger suppression at the peak probing the larger Chl~$a$ pool, with $31/25 \approx 24/18$), but systematically above the site-basis limits. Because $V_A$ is small and close to the F-2DES noise floor, we do not assign quantitative significance to this excess, nor do we combine the two peaks into a single number. The measurement confirms that essentially the entire trimer participates in incoherent mixing --- a few tens of transitions, consistent with the $\approx 18$--$24$ expected --- but does not resolve the individual manifolds precisely. A quantitative, peak-resolved $N_{\text{eff}}$, which would directly report the size of each detection manifold, is in principle accessible but requires a substantially lower F-2DES noise level.

We performed a side-by-side comparison of C-2DES and F-2DES measurements on the LHCII at 78 K, which were carried out in the same core experimental setup, using similar excitation density, but different detection schemes. The main result is that while population dynamics are clearly resolved in C-2DES, they are negligible in the F-2DES measurement. By comparing two data sets we introduced a model-independent procedure to quantify incoherent mixing directly from the data: the ratio of dynamic signal amplitudes reported in two experiments, computed at two chosen peaks. This provides a data-driven, order-of-magnitude estimate of the number of states in the system, which are connected by energy transfer pathways.
For LHCII, the measured suppression is of the order of the number of transitions in the trimer (a few tens), consistent with the $\approx 18$--$24$ expected once excitonic delocalization is taken into account. Because the suppressed dynamic signal is barely resolved above the F-2DES noise floor, we refrain from assigning a single precise value; a quantitative, peak-resolved estimate, which would report the size of each detection manifold, would require a substantially lower noise level.
Control experiments further revealed a previously unreported instrumental nonlinearity, originating in the Dazzler pulse shaper: when diffracting and shaping pulses with 100 nm bandwidth, spurious cross-peaks arise predominantly in the rephasing channel at coherence times t$_{1,3} > 100$ fs and can be reproduced in scattering measurements without a sample.
These findings underscore that action-detected cross-peaks must be interpreted with care unless incoherent mixing and shaper-induced artifacts are explicitly suppressed or subtracted.
More broadly, the results validate C-2DES as the tool of choice for tracking ultrafast population dynamics in large multichromophore systems, while F-2DES proves to be a sensitive tool for visualizing correlations between the multiple transitions in complex systems.

\section*{Acknowledgements}
The authors acknowledge A. Sahu, M. Schencker and J. Uhlig for their helpful contributions to the realization of this project.
E.A. and T.P. acknowledge financial support from the Swedish Energy Agency grant 50709-1, VR grant 2021-05207, Olle Engkvist foundation grant 235-0422, the European Union’s Horizon 2020 research and innovation program under the Marie Skłodowska-Curie grant agreement no. 945378 and the Royal Physiographic Society of Lund.
A.K., and D.Z. acknowledge funding from the Carl Trygger Foundation and Swedish Research Council.
L.B. acknowledges support through the MCIN/AEI Projects PID2021-123814OB-I00, TED2021-129241B-I00, CEX2019-000910-S, Fundacio Privada Cellex, Fundacio Privada Mir-Puig, and the Generalitat de Catalunya through the CERCA program.

\section*{Competing interests}
The authors declare no competing interests.

\section*{Author Declarations}
The manuscript reflects only the authors' view; the European Union and the Research Executive Agency are not responsible for any use that may be made of the information it contains.

\section*{Supporting information}

%

\printbibliography



  
  

\end{document}


\maketitle

\section{F-2DES setup}

\begin{figure}
\centering
\includegraphics[width=0.8\textwidth]{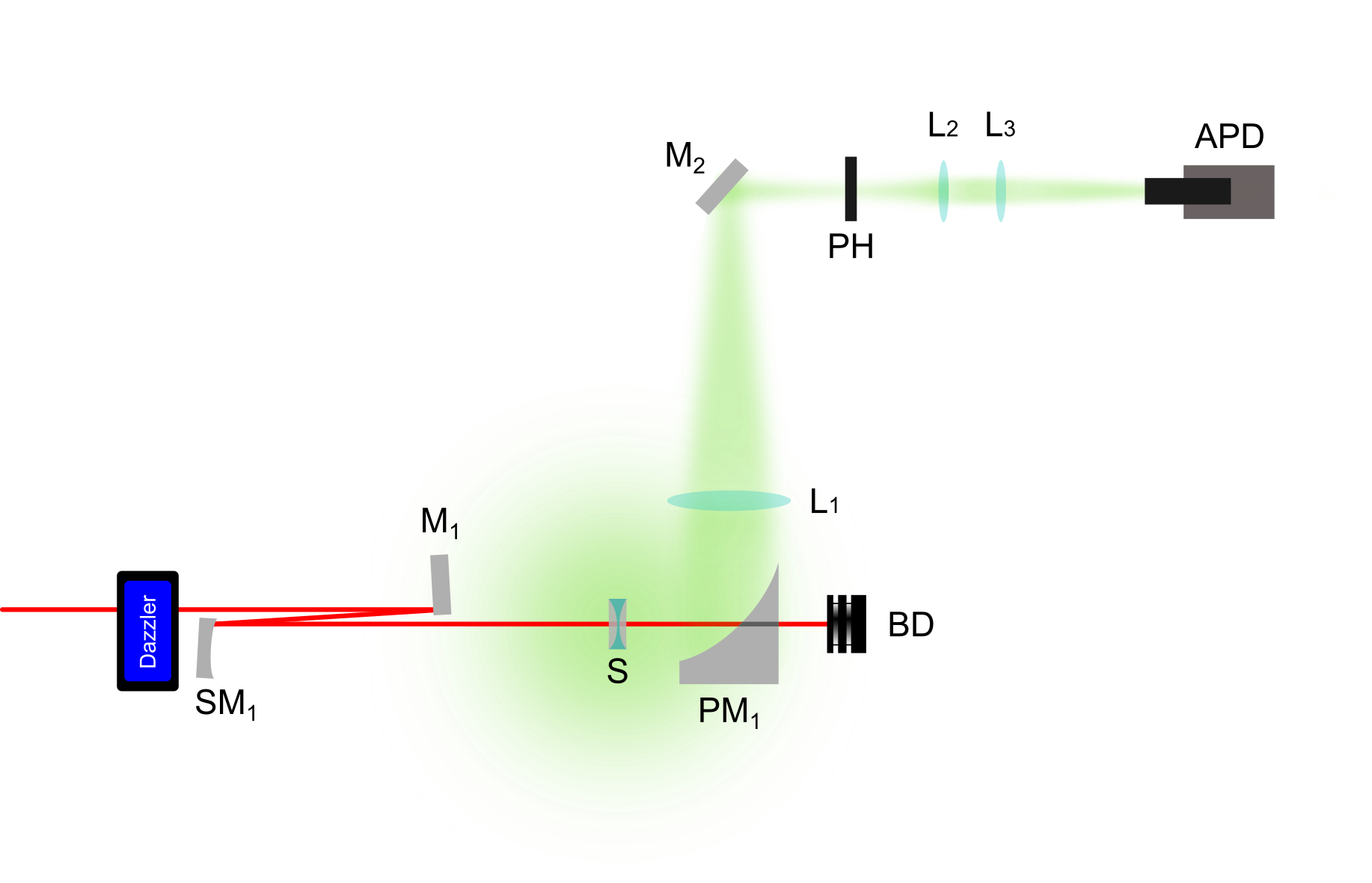}
\caption{Schematic of the experimental optical layout for pulse-shaped excitation and fluorescence detection. The excitation path (red line) utilizes an acousto-optic programmable dispersive filter (AOPDF, Dazzler) for precise pulse shaping, with the beam focused into the sample (S) by a spherical mirror (SM$_1$). The resulting fluorescence (green) is collected in a transmission geometry by an off-axis parabolic mirror (PM$_1$) specifically engineered with a central aperture. This aperture allows the primary excitation beam to pass through to a beam dump (BD) without reflection, significantly mitigating scattering contributions. The collected fluorescence is focused by lens (L$_1$) and into a pinhole (PH) to further isolate the signal from background noise. Finally, two lenses (L$_2$ and (L$_3$)) focus the filtered signal onto an avalanche photodiode (APD) for high-sensitivity detection.}
\label{fig:F2DES_Setup}
\end{figure}
\clearpage

\section{Artefact desctiption}

Notably, the population time t$_2$ beyond 300 fs diminishes the Dazzler-induced off-diagonal artifacts in the time domain. The time-domain diagonal stripes of the rephasing (R) signal evolve asymmetrically with the population time as the delay between the two pulse pairs grows.

\begin{figure}
\centering
\includegraphics[]{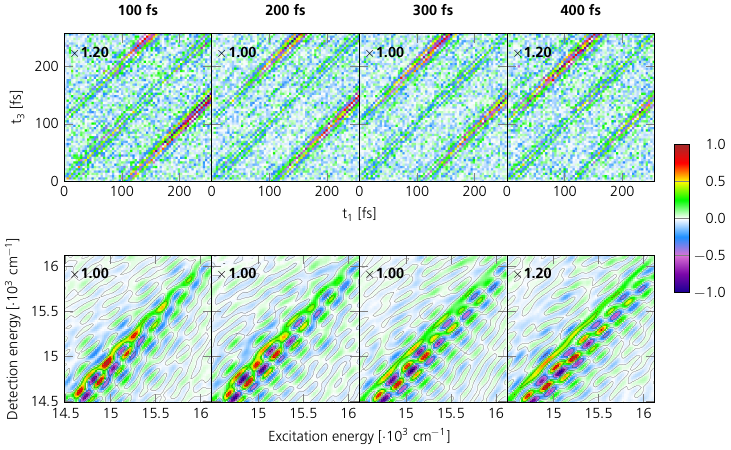}
\caption{Evolution of nonlinear artefacts in the R scattering experiment as a function of population time t$_2$. As the population time increases, the spurious nonlinear response gradually shifts toward down in antidiagonal direction.}
\label{fig:RNR2Dmaps_scatter}
\end{figure}

Limiting the coherence time scans to the initial $\sim100$ fs removes most pulse-shaper artifacts, giving significantly cleaner maps that are easier to interpret.
This comes at a predictable cost: the reduced spectral resolution, which increases from 130 cm$^{-1}$ to 330 cm$^{-1}$.
The finer features are not discernible anymore, and only two dominant peaks remain.

\begin{figure}
    \centering
    \includegraphics[]{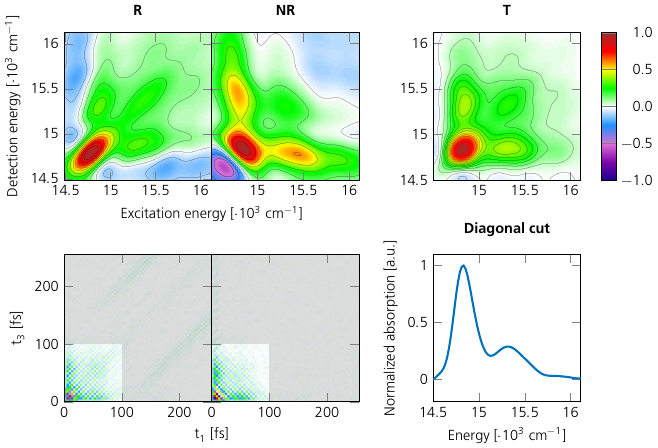}
    \caption{F-2DES maps for R, NR and T obtained by considering the coherent scans within the first 100 fs (t$_{1,3} > 100$ fs). The 2D maps in frequency domain show clean features. However, the spectral resolution is reduced to about 330 cm$^{-1}$, as illustrated by the diagonal cut of T which shows only two distinct features.}
    \label{fig:FreqTime2Dmaps_cut}
\end{figure}

\begin{figure}
    \centering
    \includegraphics[]{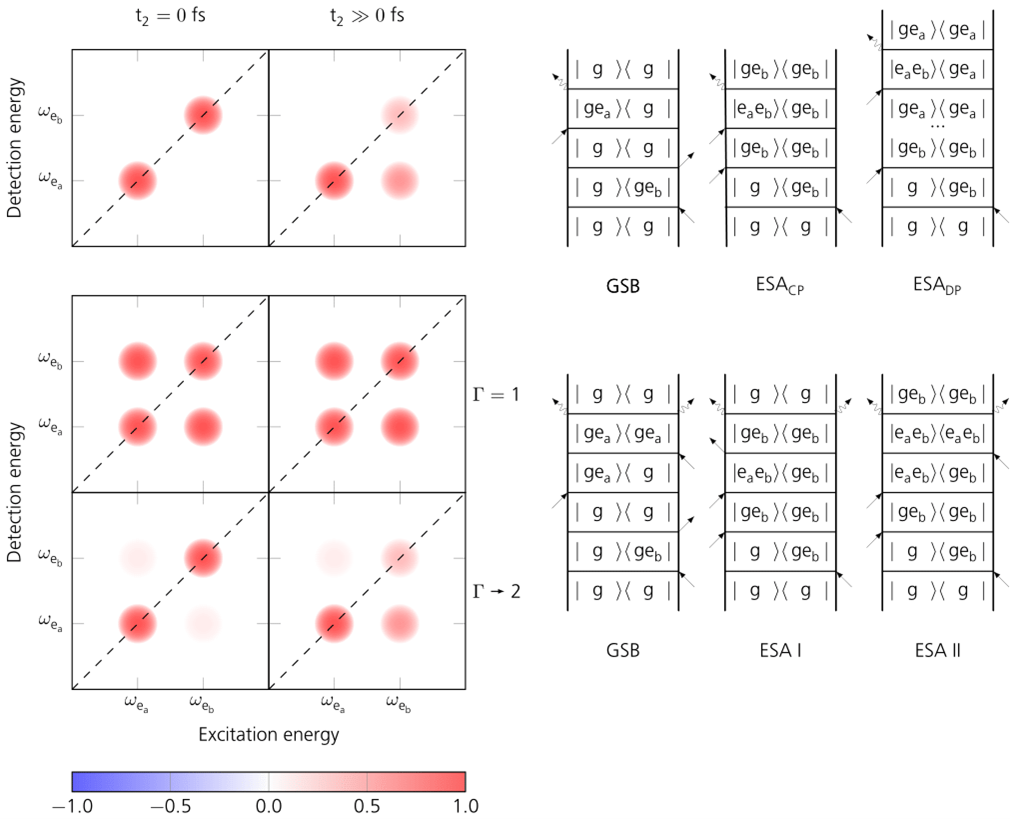}
    \caption{Schematic of R 2D spectra comparing C-2DES and F-2DES for a system with two weakly coupled 3-levels system. In C-2DES (top row) the map at t$_2 = 0$ fs shows two diagonal peaks; at long population time, the upper diagonal-peak weaken due to inner relaxation between the two coupled states from e$_\textsf{b}$ to e$_\textsf{a}$. F-2DES (bottom row) is built from incoherent mixing of excited-state populations; the mixing order $\Gamma$ characterizes the nonlinearity of the detection process. For $\Gamma$ = 1, (amount of ESA I = amount of ESA II), incoherent mixing is showing with strong ``static'' cross-peaks due to complete cancellation of ESA I and ESA II features. On the other hand, for $\Gamma$ approaching value of 2, no EEA occurs and the F-2DES starts resembling the C-2DES map.\cite{Karki2019,Kuhn2020,Bruschi2023,Bolzonello2023}}
    \label{fig:IncMix_FD}
\end{figure}
\clearpage

\section{Incoherent mixing in F-2DES of LHCII: pathway analysis and magnitude estimate}

In a coherently detected 2DES experiment, three Liouville pathways contribute to the third-order signal: ground-state bleach (GSB), stimulated emission (SE), and excited-state absorption (ESA), with relative signs $+$, $+$, $-$.
In an action-detected experiment, the fourth pulse converts the third-order coherence into a population that is read out by an incoherent observable (here, fluorescence) on the nanosecond timescale.
The fourth interaction generates two distinct ESA pathways: ESA\textsubscript{I}, which ends in a single-exciton population, and ESA\textsubscript{II}, which ends in a doubly-excited population.
The integrated signal is~\cite{PerdomoOrtiz2012,Bruschi2023}
\begin{equation}
  I = \mathrm{GSB} + \mathrm{SE} + \mathrm{ESA_I} - \Gamma\,\mathrm{ESA_{II}},
\end{equation}
where $\Gamma = \Phi_2/\Phi_1$ is the ratio of the effective quantum yields of the two- and one-exciton manifolds. 
$\Gamma = 2$ corresponds to non-interacting excitations (no annihilation; ESA pathways cancel and the F-2DES spectrum 
reduces to the coherent response with reversed sign), while $\Gamma = 1$ corresponds to complete exciton-exciton annihilation (EEA), in which the doubly-excited population produces only one unit of signal regardless of how it was created.

For a multichromophoric assembly of $N$ weakly coupled chromophores, the Liouville pathways must be further classified into \emph{self-population} pathways (all four interactions on the same site) and \emph{cross-population} pathways (two pairs of interactions on different sites). Counting these pathways yields the relations~\cite{Bolzonello2023}
\begin{equation}
\label{eq_2}
  \frac{\mathrm{SE}}{\mathrm{GSB}} = \frac{1}{N}, 
  \qquad
  \frac{\mathrm{self}}{\mathrm{cross}} = \frac{2}{N-1}.
\end{equation}
The crucial point is that excited-state population dynamics along $t_2$ are encoded \emph{only} in the SE-self pathway. 
For systems with $\Gamma \to 1$, ESA\textsubscript{I} and ESA\textsubscript{II} mutually cancel, but the cross-population GSB pathways survive and grow combinatorially with $N(N-1)$. 
The action-detected spectrum is then dominated by static cross-peaks that report on the connectivity of the assembly but obscure the population dynamics: this is the phenomenon of incoherent mixing.

For the LHCII trimer in our experimental window, all 42 chlorophylls (24 Chl~$a$ and 18 Chl~$b$) lie within the laser 
bandwidth and are connected by efficient energy transfer on timescales (sub-ps to tens of ps) much shorter than the 
fluorescence lifetime ($\sim$3--4~ns).
Excitonic delocalization within each monomer reduces the number of optically active states to roughly six to eight per monomer~\cite{Novoderezhkin2011,Renger2013}, giving an effective $N \approx 20$ across the trimer.
Equation~\ref{eq_2} then predicts an SE/GSB ratio of approximately 5\%, so that less than 5\% of the integrated F-2DES amplitude carries $t_2$ dynamics.
This is fully consistent with the essentially $t_2$-invariant cross-peaks reported in Fig.~5 of the main text 
and explains why F-2DES in this regime delivers information that is approaching that of two consecutive linear measurements rather than a true population-resolved nonlinear experiment.

We finally note two known mitigating factors~\cite{Bolzonello2023,Charvatova2025,Faitz2024}: (i) excitonic delocalization can reduce the effective $N$ below the $1/N$ bound when oscillator strength concentrates in few 
collective states (J-aggregate-like behavior), and (ii) several post-processing schemes (time gating, $T<0$ symmetry subtraction, 2Q polarization-based isolation) can suppress the GSB-cross contribution and recover the SE-self dynamics. In dense antenna complexes such as LHCII, where neither delocalization nor any of these schemes alone is sufficient, coherently detected 2DES remains the method of choice for tracking ultrafast energy flow.
\clearpage
\printbibliography